# Distortion-Based Detection of
# High Impedance Fault in Distribution Systems

Mingjie Wei, Weisheng Liu, Hengxu Zhang, Fang Shi, Weijiang Chen

*Abstract*— **Detecting the High impedance fault (HIF) in distribution systems plays an important role in power utilization safety. However, many HIFs are challenging to be identified due to their low currents and diverse characteristics. In particular, the slight nonlinearity during weak arcing processes, the distortion offset caused by the lag of heat dissipations, and the interference of background noises could lead to invalid of traditional detection algorithms. This paper proposes a distortion-based algorithm to improve the reliability of HIF detection. Firstly, the challenges brought by the diversity of HIF characteristics are illustrated with the experiments in a 10kV distribution system. Then, HIFs are classified into five types according to their characteristics. Secondly, an interval slope is defined to describe the waveform distortions of HIFs, and is extracted with the methods of the linear least square filtering (LLSF) and the Grubbs-criterion-based robust local regression smoothing (Grubbs-RLRS), so that feature descriptions under different fault conditions can be unified. Thirdly, an algorithm, as well as the criteria, is proposed to identify the fault features described by the interval slopes. Finally, the detection reliability of the algorithm is thoroughly verified with field HIF data, and the results show the improvements by comparing it to other advanced algorithms.**

## I. INTRODUCTION

HIGH impedance faults (HIFs) frequently happen on the overhead transmission lines in the medium-voltage (6-35kV) distribution systems. Generally, HIFs are caused by the accidental touches between conductors and grounding surfaces due to the broken lines or tree contacts. The high impedance materials of the grounding surfaces, like the soil, sand, asphalt, concrete, cement and tree limb, etc., will restrict the amplitudes of fault currents in the range of around 1 ampere to tens of amperes [1], making most of the protection devices invalid. According to recorded statistics [2]-[3], about 10%-20% of the faults in distribution systems are the HIFs, which will be higher if considering the unrecorded situations. Potential risks of fire hazards and human injuries [4] make it necessary and significant to enhance the reliability of HIF detections.

In most countries, more than one type of neutral exists in distribution systems, mainly including the isolated neutral, resonant neutral, and the low-resistor-earthed neutral. Therefore, after modeling the equivalent circuits of the system with different neutrals [5]-[7], many algorithms detect HIFs by expressing voltages or currents with mathematic equations, deriving the changes of amplitudes or phases before and after faults happen [8]-[10]. These model-based approaches are mathematically supported and logically demonstrated. However, they are commonly only usable in the system with specific neutrals. Meanwhile, the simplified equivalences of systems and the inherently weak features of HIFs make the detections more dependent on high-precision measurements, which is challenging both in technology and cost when further considering background noises and measuring errors. In addition, nonlinearities of HIFs also bring much trouble in the measuring accuracy.

Another group of approaches classified as the pattern recognition methodology focuses on detecting HIFs through the nonlinearity of currents or voltages [11] and has attracted wide attention since the 1970s. The nonlinearity is usually caused by AC electric arcs [1] when line conductors make poor contacts with grounding surfaces and breakdown the air. Therefore, the HIF is also called the high impedance arc/arcing fault [12]. The nonlinearities of HIFs present harmonics, waveform distortions, and sometimes are combined with unstable arcing intermittences [13]. Early works have summarized the abnormities of low-order harmonics after an HIF happens, including even-order harmonics [14], odd-order harmonics [15] and inter-harmonics [16]. Considering the complicated performances of intermittent arcs, some algorithms utilize the randomness of low-order harmonics to identify the harmonic fluctuations in this process [14], [16]. However, the harmonics caused by the noise and electronic equipment often invalidate the harmonic-based approaches, and the randomness of HIFs is usually inconspicuous for some stable arcs [13]. In recent decades, with the development of sampling techniques, researchers start taking advantage of high-frequency harmonics like the algorithms adopting the wavelet transform or some other time-frequency analyses [18]-[20], which can extract the abnormalities both during stable and intermittent arcing processes. Additionally, the waveform distortions generated by fault nonlinearities are also demonstrated useful [21]-[24]. To overcome the challenges of the criterion design due to the diversity of HIF characteristics, algorithms of artificial intelligence have been concerned about, such as the expert system [25] and the neural network [2]. However, the deficiency of representative field HIFs hinders the application of these training-based technologies.

HIF nonlinearities differ when grounding surfaces, humidity, and arcing processes are different [11]. It brings trouble for algorithms to guarantee the validity in practical applications. Generally, the arcs in the air are regarded as the primary cause of nonlinearities. However, the air arcs are not always conspicuous in HIFs, and the nonlinearities are sometimes mainly produced by

This work was supported by The National Key R&D Program of China (2017YFB0902800) and the Science and Technology Project of State Grid Corporation of China (52094017003D).

Mingjie Wei, Weisheng Liu, Hengxu Zhang and Fang Shi are all with the Key Laboratory of Power System Intelligent Dispatch and Control Ministry of Education, Shandong University, Jinan, 250061, China (e-mail: zhanghx@sdu.edu.cn).

Weijiang Chen is with the State Grid Corporation of China (e-mail: weijiang-chen@sgcc.cn).

the ionization of the solid dielectric inside the ground surface or object [23], [26]. Under the circumstances, the nonlinearity is not intense enough to produce detectable abnormalities of high-frequency harmonics. In addition, the high-frequency components are affected by the topology and can be weakened by the shunt capacitors [16] and neutral resistors [18]. As a result, many high-frequency based approaches are not valid enough in applications, especially with very low current (like less than 1A) and under the severe interferences of background noises.

In this paper, the diversity of HIF characteristics and the brought challenges are illustrated with the field faults in a 10kV distribution system. With illustrations, HIFs are classified into five types for the valid assessment of an algorithm. A distortion-based HIF detection algorithm is proposed, which is effective for various extents of nonlinearities and able to identify the offsets of waveform distortions. Moreover, the proposed algorithm is also less interfered with by background noises. In the algorithm, an interval slope is defined to describe the HIF distortions, which is extracted with the methods of linear least square filtering (LLSF) and the Grubbs-criterion-based robust local regression smoothing (Grubbs-RLRS). In this way, the HIFs grounding with different surfaces and superimposed with varying degrees of background noises can all present uniform features. It lowers the difficulties in the criterion design and improves the reliability of the fault detection.

The rest of the paper is organized as: Section II analyzes the characteristics and challenges of HIFs with the samples in a field 10kV distribution system. In Section III, a distortion-based detection algorithm is introduced. The verification of the detection reliability is carried out in Section IV by comparing it to other algorithms. Finally, conclusions are drawn in Section V.

## II. CHARACTERISTICS AND CHALLENGES

In a 10kV distribution system with the power frequency of 50Hz, a certain number of HIFs are artificially experimented (Fig.1). Measuring devices are deployed as shown in the figure and record data in the sampling frequency of 6.4kHz. In the experiments, HIFs are tested about 30 meters away from the measuring point M5 (80 meters for the electrical distance), by grounding the conductors with different surfaces materials, including dry/wet soil, dry/wet cement, dry/wet reinforced concrete, dry/wet grass, and dry asphalt concrete, etc.. Besides, three neutrals can be selected, including the isolated neutral, resonant neutral, and low-resistor-earthed neutral.

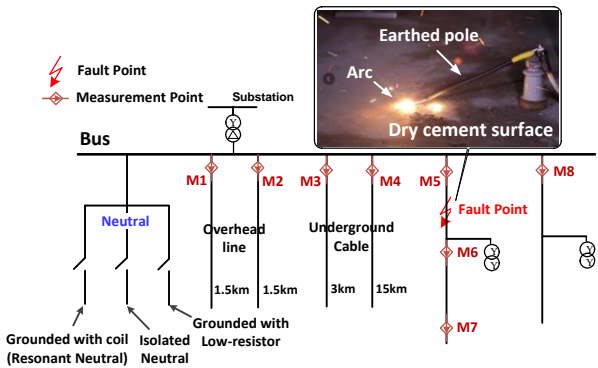

Fig.1 Topology of a 10 kV distribution system.

The nonlinearity is one of the prominent features of HIFs and is generally regarded to be produced by AC arcs. In stable situations, the nonlinearity of arc is caused by the periodic 'zero-off phenomenon' when the current is near the zero-crossing. During this period, the ionization in the arc gap is weakened due to the low current and voltage level, resulting in a decrease of the arc diameter and an increase of the arc resistance. Therefore, the current amplitudes are further limited and nonlinear distortions will exhibit. As the voltage increases after the zero-crossing, the arc ionization is enhanced, the arc resistance decreases, and the distortions of the current gradually recover [27]. As a result, distortion intervals of the fault current exist near both sides of the zero-crossing points, where the waveform distortions present the tendencies to be more parallel to the horizontal axis.

In real-world HIFs, the current distortions are affected by various factors, particularly by the grounding surface materials and air/surface humidity [11]. Except for the typical zero-off phenomenon, another three features usually presented by the practical HIF distortions need to be concerned about: 1) slight distortions in the HIFs with weak arcs; 2) distortion offsets; 3) ineffective distortions caused by background noises or arcing processes.

Firstly, the practical fault nonlinearity contains not only the impact of the air arc but also the ionization of the solid dielectric (grounding surfaces) [23], [26]. When an arc in the open air is obvious, the heat dissipation is fast, leading to the more rapid arc resistance variation and the severer current distortion (Fig.2(a)-(b)). When the arc is weak, the nonlinearity is principally caused by the ionization of the solid dielectric, where the distortion is slighter (Fig.2(c)-(e)).

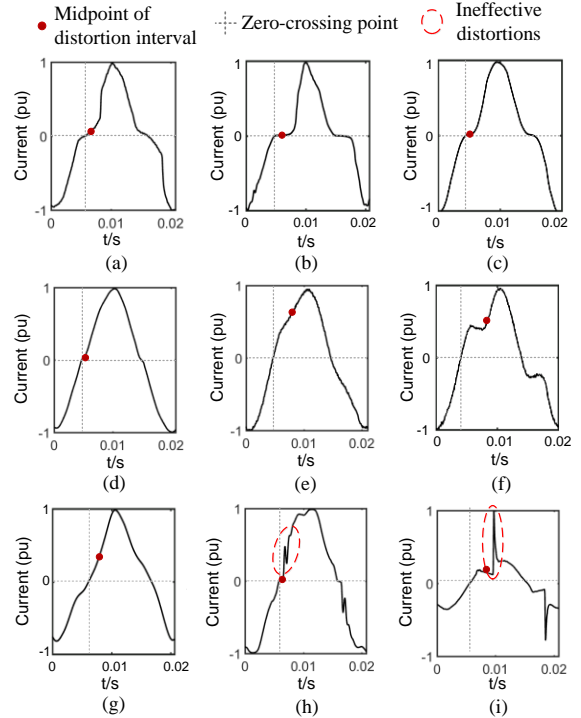

Fig.2 Current waveform of the HIFs tested in the 10 kV distribution system: (a) wet asphalt concrete, isolated neutral; (b) wet soil, low-resistor-earthed neutral; (c) dry soil, isolated neutral; (d) wet cement, low-resistor-earthed

neutral; (e) dry grass, resonant neutral; (f) wet grass, resonant neutral; (g) dry cement, resonant neutral; (h) dry cement pole, isolated neutral; (i) dry soil, resonant neutral;

The slight distortions are commonly smoother and cannot generate many high-frequency abnormalities. Therefore, many high-frequency-based algorithms that have been widely researched in recent decades [18]-[20] will be invalid. For example, Fig.3 exhibits a comparison between the two field HIFs, and the high-frequency components from 1 kHz to 5 kHz are extracted with the wavelet transform. Raw waveforms are both preprocessed by the wavelet filter. For the HIF in Fig.3(a) where the arc is obvious and presents severer distortions, the high-frequency components show considerable abnormalities compared to the pre-fault state (before 0 seconds). In these cases, the arc can be well detected. Nevertheless, for the HIFs with slighter distortions, the high-frequency components are significantly weakened, failing to distinguish from pre-fault conditions (Fig.3(b)). Moreover, the slighter HIF distortions are usually accompanied by the lower currents, which make high-frequency components easier to be covered by noises.

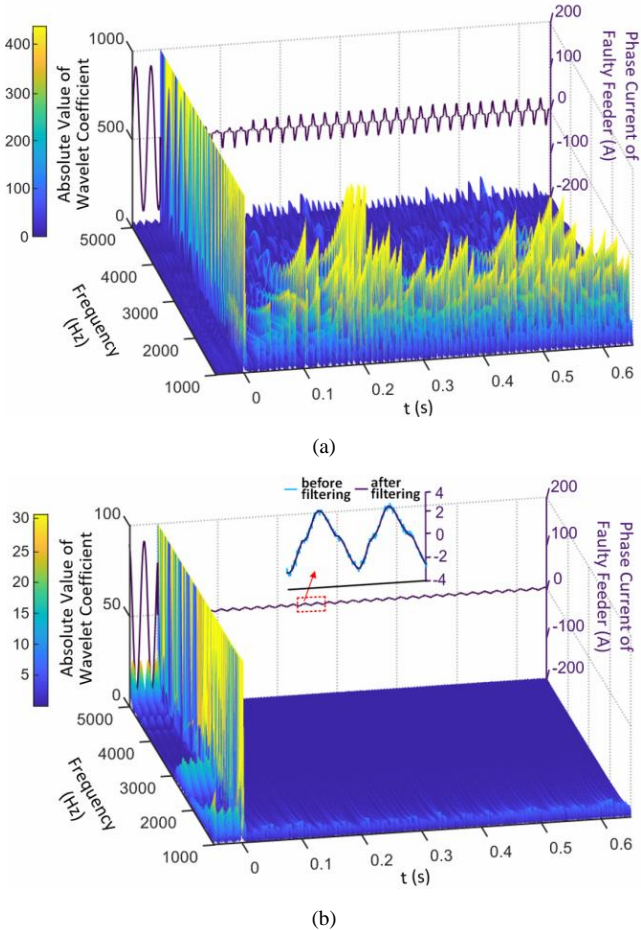

(a)

(b)

Fig.3 Abnormalities in the high-frequency region of HIFs grounding with different surfaces in a resonant grounding system (fault happens at 0s): (a) wet soil, (b) dry asphalt concrete.

Secondly, in ideal conditions, the current distortion caused by the 'zero-off phenomenon' should be near and symmetric to the zero-crossing point. Practically, due to the differences in the heat dissipation capabilities of the ionization space, the dissipated power lags behind the energizing power to varying degrees, leading to the offsets of distortions relative to the zero-crossing points, especially in Fig.2(e)-(f) (see the relative position between the zero-crossing point and the mid-point of the distortion interval). The neglect of distortion offsets could cause the malfunctions of detection algorithms. For example, [23] proposes a voltage-current characteristics profile (VCCP) to describe the nonlinear distortions of HIFs. For the HIFs with small distortion offsets, no matter with obvious or weak arcs, the VCCPs typically exhibit good hysteresis loops as shown in Fig.4(a). They are all with large slopes near the origin and much smaller ones near the extremums. However, for the HIFs with large distortion offsets, the VCCPs may not present the above phenomena (Fig.4(b)) but show two vastly different trajectories.

Thirdly, the ineffective distortions caused by arcing processes or background noises also bring troubles in accurately extracting the effective distortions, i.e., the 'zero-off phenomenon'. Generally, the background noise is composed of natural electromagnetic interferences, measuring errors, and the harmonics of electrical equipment. Fig.4(c) also takes examples with VCCPs, which show more complicated trajectories compared to the typical conditions.

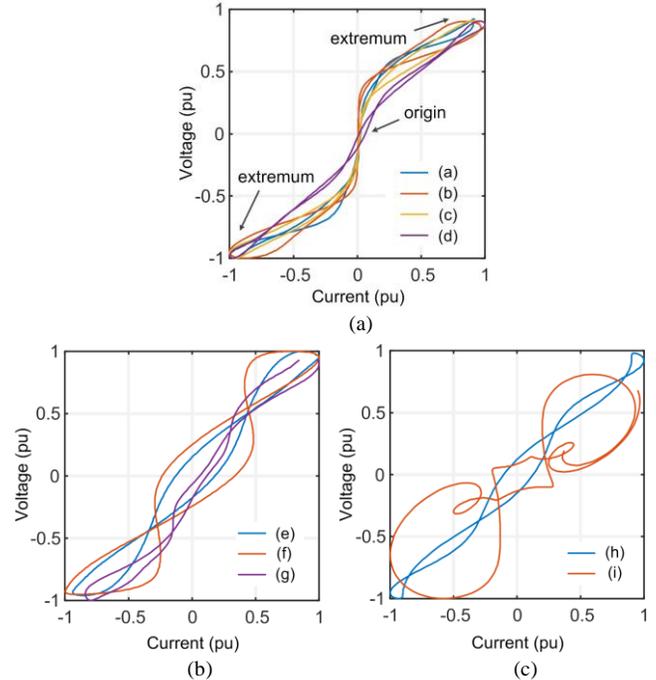

(a)

(b)                    (c)

Fig.4 VCCPs for different HIFs in Fig.2 (after filtering), including the VCCPs of (a) typical HIFs, (b) HIFs with large distortion offsets, and (c) HIFs with large ineffective distortions.

As a result, an HIF generally has one or several of the following features:

1) with obvious arcs and severe distortions, like the HIFs in Fig.2(a), (b), (f), (h) and (i);

2) with weak arcs and slight distortions, like the HIFs in Fig.2(c)-(e) and (g);

3) with large distortion offsets, like the HIFs in Fig.2(e)-(g) and (i).

4) with ineffective distortions, such as the impulse noises

caused by arcing processes, like the HIFs in Fig.2(h) and (i).

With the summarized features, a classification is made as follows to better assess the validities of algorithms in detecting different types of HIFs:

Type A is with feature 1) but without feature 2)-4);
Type B is with feature 2) but without feature 1), 3) and 4);
Type C1 is with feature 1), 3) but without feature 2), 4):
Type C2 is with feature 2), 3) but without feature 1), 4);
Type D is with feature 4).

## III. DETECTION OF HIF

### A. Description of the Interval Slope

The fault current or the zero-sequence current in an HIF is much smaller than the rated value and thereby can be more interfered with by the background noise. As a result, a detection algorithm with a good anti-noise ability is primary to guarantee the feature extraction and description.

Firstly, zero-sequence currents need the preprocessing by a wavelet filter (WF) or low-pass filter (LPF). The WF carries out de-noise in various frequency bands, while the LPF eliminates the components over the cut-off frequency. If using LPF, the cut-off frequency cannot be too low to avoid enlarging transient interferences, especially for the impulse noise that frequently happens in the arcing process. The cut-off frequency is suggested to set around 1500Hz.

Secondly, waveform distortions of zero-sequence currents can be reflected by the derivative. However, to restrain the fluctuation of the derivative due to background noises, an interval slope is defined to describe the distortions by using the linear least-square fit (LLSF) and the Grubbs-criterion-based robust local regression smoothing (Grubbs-RLRS). It is introduced in detail as follows:

For a zero-sequence current $i_0(n)$, its interval slope at the sampling point $n_s$ is denoted as $IS_{i_0}(n_s)$, which is described with the LLSF. Its absolute value $\left| IS_{i_0}(n_s) \right|$ is expressed as:

$$\left| IS_{i_0}(n_s) \right| = \left| \frac{l \cdot \sum_{INT_{n_s}} [n \cdot i_0(n)] - \sum_{INT_{n_s}} n \cdot \sum_{INT_{n_s}} i_0(n)}{l \cdot \sum_{INT_{n_s}} n^2 - \left( \sum_{INT_{n_s}} n \right)^2} \right| \quad (1)$$

$$INT_{n_s}: n_s - \frac{l}{2} < n \le n_s + \frac{l}{2}$$

where, the interval $INT_{n_s}$ is with the length of $l$ and lets $n_s$ as the midpoint; $l$ is suggested as $N_T/8$ and $N_T$ represents the number of sampling points in a power frequency cycle. In the following paper, the interval slope represents the absolute value $\left| IS_{i_0}(n_s) \right|$ unless specially indicated.

For a sinusoidal waveform, the interval slope decreases to the minimum only near the minimal or maximal instantaneous currents as shown in Fig.5(a), which presents the shape of a 'double Λ' in a cycle. However, for a distorted waveform in Fig.5(b), the interval slope also decreases to the minimum near the distortion intervals, which presents the shape of 'double M'. Therefore, features of interval slopes can be used to identify HIFs and distinguish from the non-fault conditions.

However, to eliminate the impacts of ineffective distortions,

especially the impulse noises or some other irregular distortions, a method of Grubbs-RLRS is used to filter the zero-sequence current before calculating the interval slopes.

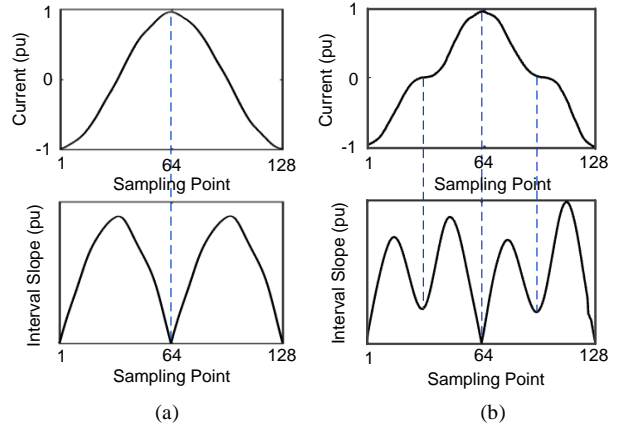

Fig.5 Interval slopes in (a) non-fault and (b) fault conditions.

For $n_s$, define the $i_0(n)$ that belongs to the interval $INT_{n_s}$ as $i_0^{INT_{n_s}}(n)$. Then, the RLRS is used to build a $m$-order polynomial $f_m^{INT_{n_s}}(n) = \alpha_0 + \alpha_1 n + \cdots + \alpha_m n^m = \sum_{j=0}^{m} \alpha_j n^j$ to fit the $i_0^{INT_{n_s}}(n)$. The fitting error is expressed as:

$$\xi = \sum_{i=1}^{l} \varepsilon_i = \sum_{i=1}^{l} \left[ i_0^{INT_{n_s}}(n_i) - \sum_{j=0}^{m} \alpha_j n_i^j \right]^2 \cdot w_i \quad (2)$$

where, $w_i$ is the weight coefficient initiated as 1. The purpose of RLRS is to find out a $\boldsymbol{\alpha} = \{\alpha_j | j = 0,1, \dots, m\}$ that makes $\xi$ become the minimum. Let $\xi' = 0$ and the $\boldsymbol{\alpha}$ can be calculated as follows [28]:

$$\boldsymbol{\alpha} = (\boldsymbol{N}^T \boldsymbol{W} \boldsymbol{N})^{-1} \boldsymbol{N}^T \boldsymbol{W} \boldsymbol{I} \quad (3)$$

where, $\boldsymbol{N} \in \mathbb{R}^{l \times (m+1)}$ and $\boldsymbol{N}_{i,j} = n_i^j$; $\boldsymbol{W} \in \mathbb{R}^{l \times l}$, which is a diagonal matrix and $\boldsymbol{W}_{i,i} = w_i$ ; $\boldsymbol{I} \in \mathbb{R}^{l \times 1}$ and $\boldsymbol{I}_{i,1} = i_0^{INT_{n_s}}(n_i)$.

Ineffective distortions like the impulse signals show fast variation and recovery compared to the distortions of the 'zero-off phenomenon'. To eliminate the impacts of the ineffective distortions on the feature extraction, $w_i$ and $\boldsymbol{\alpha}$ are updated based on the Grubbs-criterion [29], which excludes the abnormal values with the idea of iterative screening and has been widely used for its better performances in outlier screening than other same types of algorithms. The normalized residual of the Grubbs-criterion is defined as:

$$G_i = \frac{\varepsilon_i - \overline{\boldsymbol{\varepsilon}}}{STD(\boldsymbol{\varepsilon})} \quad (4)$$

where, only the $\varepsilon_i$ with $w_i = 1$ is considered in the calculation, i.e., $\boldsymbol{\varepsilon} = \{\varepsilon_i \mid w_i = 1 \text{ and } i = 0,1, \dots, m\}$. $\overline{\boldsymbol{\varepsilon}}$ represents the average value of $\boldsymbol{\varepsilon}$, and $STD(\boldsymbol{\varepsilon})$ represents its standard deviation. Update the $w_i$ as follows:

$$\begin{cases} w_i = 1, \ G_i < G_{p,N} \\ w_i = 0, \ G_i \ge G_{p,N} \end{cases} \quad (5)$$

where, $G_{p,N}$ represents the Grubbs threshold with confidence probability of $p$ generally ranging in 90%~99.5%. $p$ is set as 90% in the paper. The value of $G_{p,i}$ with different $p$ and $N$ can be referred to TABLE II in the Appendix. After the update of $w_i$, (2)-(5) are conducted circularly until no $w_i$ is updated from 1 to 0 anymore. Then, $\left|IS_{i_0}(n_s)\right|$ is calculated by replacing the original $i_0(n)$ with the $i_0^{INT_{n_s}}(n)$.

Finally, a curve of the interval slope can be achieved by calculating the $\left|IS_{i_0}(n)\right|$ point by point. However, $\left|IS_{i_0}(n)\right|$ can also be calculated every a few points and then uses the interpolation to accelerate the calculation. The effectiveness of

the Grubbs-RLRS is illustrated in Fig.6 (a)-(c). In Fig.6(a), it shows that the ineffective impulse noise will be enlarged if the cut-off frequency $f_c$ of the LPF is too low, which apparently makes the impact of the ineffective distortion more difficult to be eliminated. Fig.6(b) presents the derivative of the zero-sequence current, compared with which, the interval slope described by LLSF in Fig.6(c) can better restrain the fluctuations caused by the arcing process. Moreover, Fig.6(c) also demonstrates the advantages of the Grubbs-RLRS method in eliminating the interference of the impulse noise, which guarantees the correct presentation of the 'double M'.

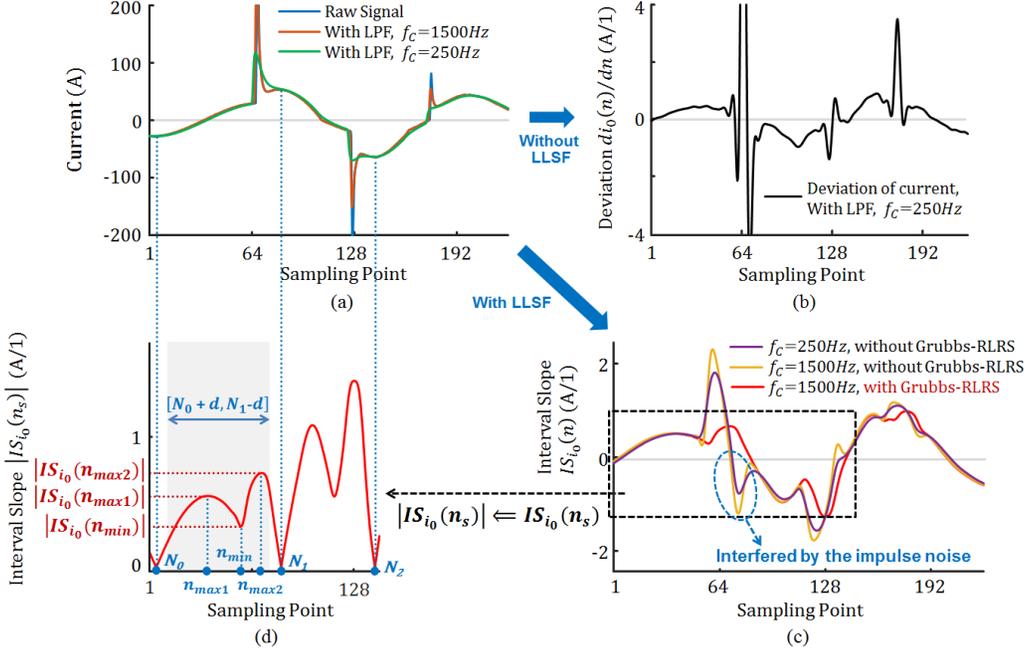

Fig.6 Effectiveness of the LLSF and Grubbs-RLRS in the description of interval slopes: (a) The zero-sequence current of a field HIF shown in Fig.2(i), $f_c$ represents the cut-off frequency of the LPF; (b) The curve of the derivative; (c) The curve of the interval slope (not the absolute value) after utilizing the LLSF with or without the Grubbs-RLRS; (d) The schematic diagram to detect distortions with the interval slope (the absolute value).

### B. Judgment

After the description of interval slopes, the distorted features of HIFs can be detected by the following procedures:

Step 1: As shown in Fig.6(d), in each cycle, find out the positions of the maximal and minimal zero-sequence currents with the fast Fourier transform (FFT). The abscissas of the two positions are denoted as $N_1$ and $N_2$, respectively. Considering the FFT deviation, $N_1$ and $N_2$ are further calibrated as the minimal interval slopes in the vicinity. $N_0$ represents the minimal interval slope in the vicinity of $N_1 - N_T/2$.

Step 2: For the half-cycle interval $[N_0, N_1]$, $n_{min}$ is a minimal point in $[N_0 + d, N_1 - d]$, i.e., $\left|IS_{i_0}(n_{min} - 1)\right| \leq \left|IS_{i_0}(n_{min})\right| \leq \left|IS_{i_0}(n_{min} + 1)\right|$ as shown in Fig.6(d), where the short zone $2d$ respectively in the vicinity of $N_0$ and $N_1$ is to neglect the $n_{min}$ near the maximal or minimal instantaneous zero-sequence currents. Only when the following two criteria are satisfied, the half-cycle interval $[N_0, N_1]$ is claimed to possess the characteristic of the HIF nonlinearity:

1) The first $n_{min}$ is denoted as $n_{min0}$ and should satisfy:

$$\begin{cases} \left|IS_{i_0}(n_{min})\right| \leq K_{set1} \cdot \overline{\left|IS_{i_0,nmax}\right|} \\ Num_{c1} = Num_{c2} = 2 \end{cases} \quad (6)$$

where, $\overline{\left|IS_{i_0,nmax}\right|} = \left(\left|IS_{i_0}(n_{max1})\right| + \left|IS_{i_0}(n_{max2})\right|\right)/2$, and $\left|IS_{i_0}(n_{max1})\right|$, $\left|IS_{i_0}(n_{max2})\right|$ are two maximums respectively in the interval of $(N_0, n_{min})$ and $(n_{min}, N_1)$ as shown in Fig.6(d); $K_{set1}$ represents the sensitivity coefficient and is generally set as 0.80~0.85 in order to guarantee the detection reliability and security at the same time (the statistical basis of the threshold is shown in Fig.11 of the Appendix); $Num_{c1}$ and $Num_{c2}$ are the numbers of the sampling points $n_{c1}$ and $n_{c2}$ that satisfy (7) and (8), respectively:

$$\left|IS_{i_0}(n_{c1})\right| = \frac{\left|IS_{i_0}(n_{min})\right| + \left|IS_{i_0}(n_{max1})\right|}{2}, n_{c1} \in (N_0, n_{min}) \quad (7)$$

$$\left|IS_{i_0}(n_{c2})\right| = \frac{\left|IS_{i_0}(n_{min})\right| + \left|IS_{i_0}(n_{max2})\right|}{2}, n_{c2} \in (n_{min}, N_1) \quad (8)$$

2) Due to the filtering residues of the Grubbs-RLRS, there could be more than one $n_{min}$ in $[N_0 + d, N_1 - d]$. If so,

except for $n_{min0}$, all the other $n_{min}$ should satisfy:

$$\begin{cases} n_{min} \in (n_{max1}, n_{max2}) \\ \dfrac{\left| IS_{i_o,nmax} \right| - \left| IS_{i_o}(n_{min}) \right|}{\left| IS_{i_o,nmax} \right| - \left| IS_{i_o}(n_{min0}) \right|} \in [1 - K_{set2}, 1 + K_{set2}] \end{cases} \quad (9)$$

where, $K_{set2}$ also represents the sensitivity coefficient. $K_{set2}$ is set as 0.05~0.25 (also based on statistical analyses).

Step 3: In the half-cycle interval of $[N_1, N_2]$, carry out the same identification as the $[N_0, N_1]$. If both the two half-cycle intervals possess the characteristics of the HIF nonlinearity, this cycle is recorded as a 'faulty cycle'.

Step 4: Pursue the judgments of Step 1-3 cycle by cycle. If there are a few successive cycles recorded as the 'faulty cycle', the HIF is detected. Empirically, the number of the successive cycles to confirm an HIF is set as 4~6.

## IV. CASE STUDY

The detection reliability requires an algorithm to be able to identify various HIF distortions and have good immunity to the probable background noises. In this section, 28 field HIFs in the 10kV distribution system introduced in Section II are used for verifications.

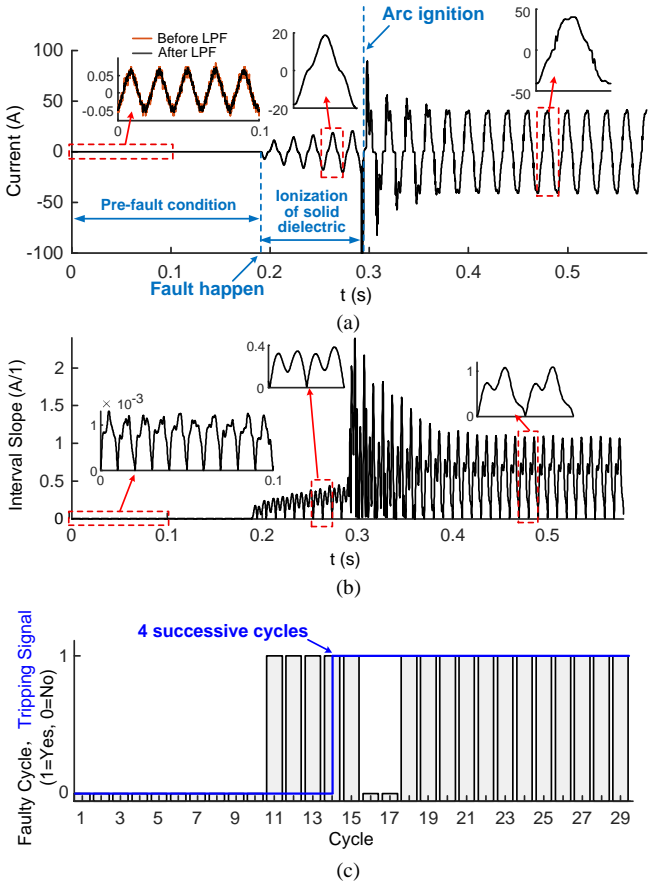

Fig.7 Detection result of a field HIF in Fig.2(h): (a) Zero sequence current; (b) Interval slope; (c) Judgment of the 'faulty cycle' and performance of the tripping signal.

Set $K_{set1}$=0.83, $K_{set2}$=0.1, and the detection result of the HIF in Fig.2(h) is illustrated in Fig.7, where the fault happens around 0.19s. During 0.19s~0.29s, the nonlinear distortions are mainly produced by the ionization of the solid dielectric, whereas the arc ignition occurs at about 0.29s and exerts dominating influences on the waveform distortions after that. Whichever physic process produces the distortions, the 'double M' of interval slopes can be reliably identified as shown in Fig.7(b). In addition, before the fault happens at 0.19s, distortions also exist due to the severe background noises caused by measuring errors. However, it can be correctly distinguished by the proposed criteria as shown in Fig.7(c).

Compared to the VCCPs in Fig.4, the proposed distortion-based algorithm can unify the features of all types of HIFs classified in Section II, which all present interval slopes with the pure 'double M' in each cycle (Fig.8) by combining the methods of LLSF and Grubbs-RLRS.

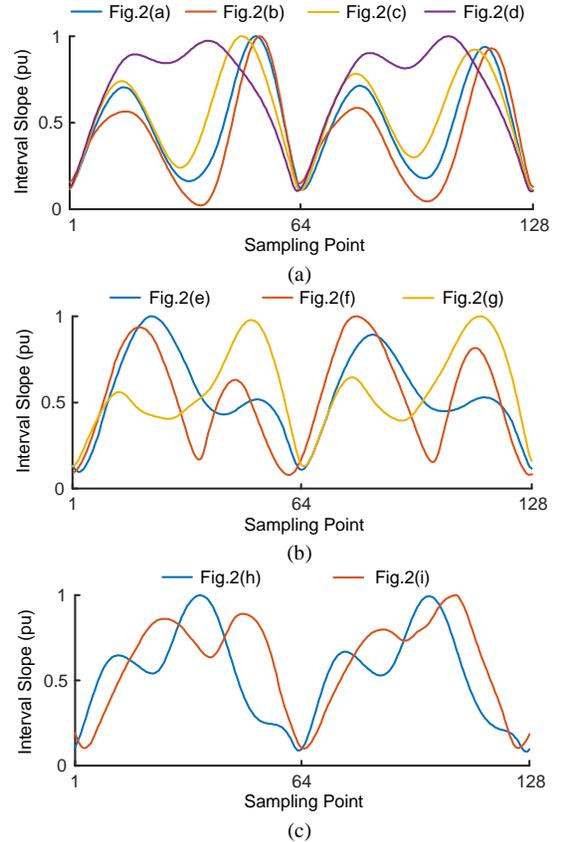

Fig.8 Interval slopes of different types of HIFs (illustrated by the HIFs in Fig.2): HIFs of (a) Type A and B, (b) Type C1 and C2, and (c) Type D.

Then, with the proposed criteria, the 'double M' features can all be reliably identified. In the 28 field HIFs, there contain 5 Type A, 8 Type B, 3 Type C1, 6 Type C2, and 6 Type D HIFs. Including the comparisons with another three algorithms, the detection results of these different types of HIFs are shown in TABLE I. Specifically, the algorithm in [20] detects HIFs based on the high-frequency abnormalities, whereas the algorithms in [23] and [24] detect HIFs also with waveform distortions. In [20], an HIF is detected by identifying the fast rising of an energy defined by the discrete wavelet transform (DWT). As shown in TABLE I, this high-frequency-based algorithm can well detect the HIFs classified as Type A, C1 and D, but is invalid for the HIFs of Type B and C2 because of their slight distortions as explained in Fig.3 of Section II. [23] describes the distortions of HIFs with the VCCPs introduced in Fig.4, which performs well in detecting the HIFs of Type A, B and C1 but cannot completely guarantee the reliability, especially for the HIFs of Type C2. [24] proposes a method to describe the distortion by its convex and concave characteristics (CCC), which presents difficulties in figuring out the large distortion offsets, i.e., the HIFs of Type C1, C2, and ineffective distortions, i.e., the HIFs of Type D. In contrast, with the processing of the LLSF and Grubbs-RLRS, as well as the effective criteria, the proposed algorithm can correctly detect all of these samples.

TABLE I NUMBER OF HIFs DETECTED BY DIFFERENT ALGORITHMS

| Method | Detected Number of Different HIF Types | | | | | Detected Num (Rate) |
|---|---|---|---|---|---|---|
| | A | B | C1 | C2 | D | |
| DWT | 5 | 1 | 3 | 0 | 6 | 15 (53.57%) |
| VCCP | 5 | 8 | 3 | 2 | 5 | 22 (78.57%) |
| CCC | 5 | 8 | 1 | 0 | 3 | 17 (60.71%) |
| LLSF-Grubbs-RLRS | 5 | 8 | 3 | 6 | 6 | 28 (100.0%) |
| Total | 5 | 8 | 3 | 6 | 6 | |

Besides, the anti-noise ability is also significant to guarantee the reliability of the detection algorithm in the distribution system, which is usually with a noisy environment in practical situations. Also using the 28 field HIFs, Fig.9 shows the detected rate comparisons of the above four algorithms under different noise levels, where the noise is composed of the practical noise and the complementary simulated noise. It demonstrates that the proposed algorithm has a stronger capability of noise immunity, as well as the algorithm of VCCP. They can both guarantee the reliability when the signal-to-noise ratio (SNR) is above 16dB and becomes increasingly invalid when the SNR is below 5dB. The detailed detection result of a field HIF with intense background noises is shown in Fig.10. However, for the algorithm of DWT, it can be significantly interfered with by noises and is completely invalid when the SNR is below 30dB.

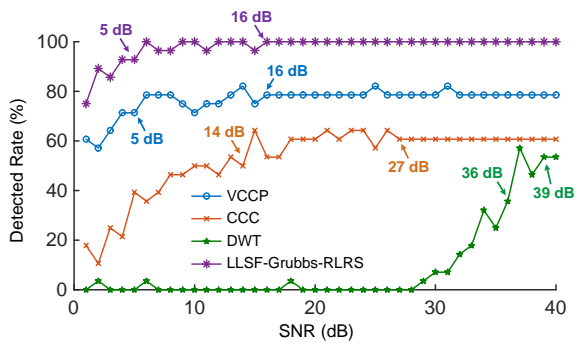

Fig.9 Detected rate of different algorithms under various noise levels.

The security of a detection algorithm is equally essential for the practical application, which requires the algorithm not to send out the tripping signal under non-fault conditions. Apparently, the typical switching operations in the system, like the switchings of capacitors or loads, cannot generate the distortions with the 'double M' interval slopes. Besides, the nonlinear loads at the low-voltage side also cannot affect the zero-sequence current at the medium-voltage side because of the filtering of the step-down transformer, whose wirings at the primary side are generally with unearthed neutral [13], [23], [30]. In addition, it is also illustrated in Fig.7, Fig.10 and Fig.11(c) that the intense background noises during the non-fault conditions do not lead to misjudgments. As a result, the security of the HIF detection can be guaranteed by the proposed distortion-based algorithm.

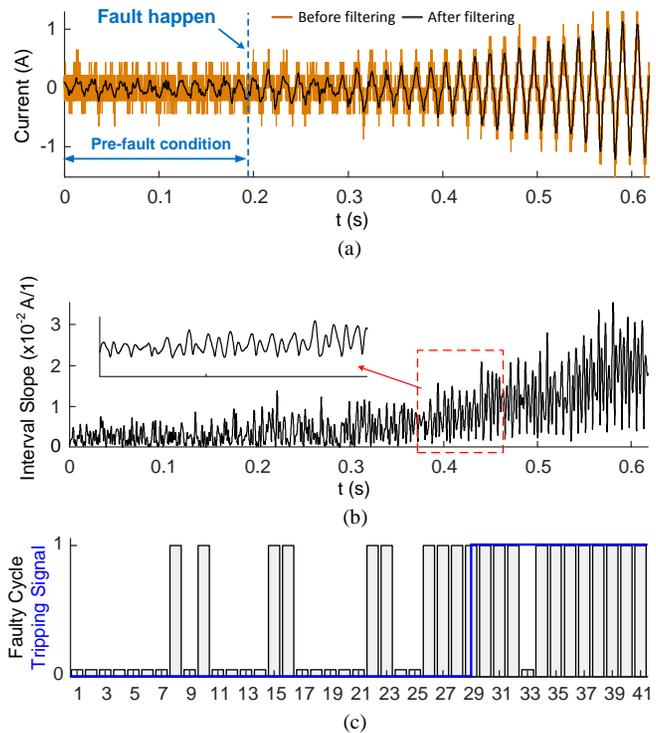

Fig.10 Detection result of a field HIF under the severely noisy environment (average 10.63dB): (a) Zero sequence current; (b) Interval slope; (c) Judgment of the 'faulty cycle' and performance of the tripping signal.

## V. CONCLUSION

The HIF detection in the distribution system is a challenging and significant subject. To better understand the diversity presented by the characteristics of HIFs, this paper classified HIFs into five types. For many existing algorithms, only two to three types can be effectively identified. Therefore, this classification can help better assess the reliability of an algorithm when the diversity of fault features is under consideration. A distortion-based algorithm is proposed in the paper, describing the features with the defined interval slopes, which are extracted by the methods of LLSF and Grubbs-RLRS. Their combination shows good effectiveness in correctly presenting the 'double M' characteristics of interval slopes. Compared to an advanced high-frequency-harmonic-based algorithm and another two distortion-based algorithms, the proposed algorithm shows its superiorities in detecting all these five types of HIFs, as well as the better immunity to background noises.

# APPENDIX

TABLE II GRUBBS THRESHOLD $G_{p,N}$

| $p$ $N$ | 90.0% | 95% | 97.5% | 99.0% | 99.5% |
|---|---|---|---|---|---|
| 5 | 1.602 | 1.672 | 1.715 | 1.749 | 1.764 |
| 6 | 1.729 | 1.822 | 1.887 | 1.944 | 1.973 |
| 7 | 1.828 | 1.938 | 2.020 | 2.097 | 2.139 |
| 8 | 1.909 | 2.032 | 2.126 | 2.22 | 2.274 |
| 9 | 1.977 | 2.110 | 2.215 | 2.323 | 2.387 |
| 10 | 2.036 | 2.176 | 2.290 | 2.410 | 2.482 |
| 11 | 2.088 | 2.234 | 2.355 | 2.485 | 2.564 |
| 12 | 2.134 | 2.285 | 2.412 | 2.550 | 2.636 |
| 13 | 2.175 | 2.331 | 2.462 | 2.607 | 2.699 |
| 14 | 2.213 | 2.371 | 2.507 | 2.659 | 2.755 |
| 15 | 2.247 | 2.409 | 2.549 | 2.705 | 2.806 |
| 16 | 2.279 | 2.443 | 2.585 | 2.747 | 2.852 |
| ... | ... | ... | ... | ... | ... |

$p$ represents confidence probability and reflects the strict degree of abnormality screening. Abnormalities are more easily to be eliminated as $p$ decreases. $N$ represents the number of samples in the sequence for screening. The sampling frequencies of the field and simulated HIF in the paper are both 6.4kHz, i.e. $N \le 16$ when the length of interval slope is set as $l = N_T/8$.

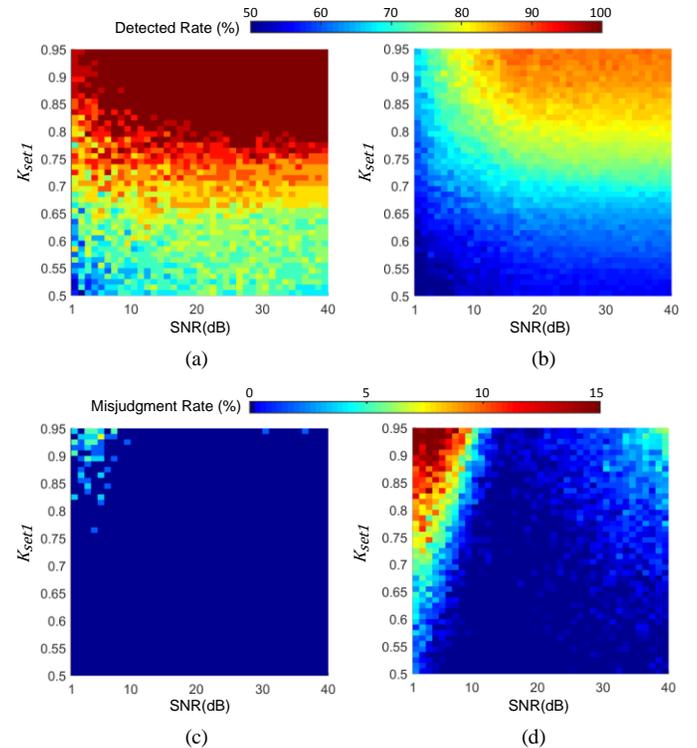

Fig.11 Statistical analysis of the threshold $K_{set1}$ when noise levels are different ($K_{set2} = 0.1$): (a) Detected rate of 28 HIFs; (b) Detected rate of 572 faulty cycles in the 28 HIFs; (c) Misjudgment rate of the non-fault conditions in the 28 HIFs; (d). Misjudgment rate of 504 non-fault cycles in the 28 HIFs.